%% file: main.tex
\documentclass[journal]{IEEEtran}
\pdfoutput=1

\usepackage{amsmath,amsfonts}
\usepackage{multirow}
\usepackage{graphicx}
\usepackage{textcomp}
\usepackage{xcolor}
\usepackage{framed}
\usepackage{listings}
\usepackage{graphicx,subfigure}
\usepackage{bbding}
\usepackage{color}
\usepackage{makecell}
\usepackage{enumitem}
\usepackage{xspace}
\usepackage{threeparttable}
\usepackage{verbatim}
\usepackage{hyperref}
\usepackage{pifont}
\usepackage{cleveref}
\usepackage{colortbl}
\usepackage{booktabs}
\usepackage{amsmath}
\usepackage{xspace}
\usepackage{epstopdf}
\usepackage[vlined, ruled,linesnumbered]{algorithm2e}
\usepackage[font=small,skip=2pt]{caption}

\def\eg{\emph{e.g.}} 
\def\ie{\emph{i.e.}}

\newcommand{\distance}{2pt}
\AtBeginDocument{%
  }

\begin{document}



\title{Benchmarking and Revisiting Code Generation Assessment: A Mutation-Based Approach}

\author{Longtian Wang,
\thanks{Longtian Wang, Yuhan Zhi, and Chao Shen are with the School of Cyber Science and Engineering, Xi'an Jiaotong University, Xi'an 710049, China (e-mail: chaoshen@mail.xjtu.edu.cn).}
        Tianlin Li,
        \thanks{Tianlin Li is with the Nanyang Technological University, Singapore 639798.}
        Xiaofei Xie,
        \thanks{Xiaofei Xie and Jian Wang is with Singapore Management University, Singapore 188065 (e-mail: xfxie@smu.edu.sg).}
        
        Yuhan Zhi,
        Jian Wang,
        Chao Shen
        
        }




\maketitle

\input{section/0-abstract}


\IEEEpeerreviewmaketitle

\input{section/1-Introduction}

\input{section/2-Background_and_Related_Work}

\input{section/2.5-Preliminary}

\input{section/3-Methodology}

\input{section/4-Experiments}

\input{section/5-Discussion}
\input{section/6-Conclusion}



\ifCLASSOPTIONcaptionsoff
  \newpage
\fi

\bibliographystyle{IEEEtran}
\bibliography{sample-base}

\end{document}

%% file: section/0-Abstract.tex
\begin{abstract}

Code Large Language Models (CLLMs) have exhibited outstanding performance in program synthesis, attracting the focus of the research community. The evaluation of CLLM's program synthesis capability has generally relied on manually curated benchmarks.
However, there is a substantial gap between real-world
scenarios and benchmark settings. Existing benchmarks typically provide only a single input prompt for the evaluation of each synthesis problem. However, in practice, a problem can be described in various ways, including with typos, where developers may struggle to understand certain descriptions and seek clarification to find more suitable wording. Such various descriptions may lead to variations in the performance of CLLMs on the same question, resulting in a biased evaluation when using existing benchmarks. 
In this paper, we aim to explore these pitfalls with the goal of revisiting and enhancing future benchmark designs. To simulate real-world variations in problem descriptions, we propose 10 mutation strategies and introduce three new metrics to evaluate their impact on code generation. We then assess five popular CLLMs using 12,834 generated prompt variants, and found a significant performance discrepancy between the results from existing benchmarks and those from mutated benchmarks containing perturbations and variations. This finding underscores the need for more robust evaluation methods and benchmarks.





\end{abstract}

%% file: section/1-Introduction.tex
\section{Introduction} \label{introduction}

Code Large Language Models (CLLMs), which are constructed on the architecture of large language models and trained on vast corpora of source code,  have demonstrated remarkable performance in various code-related tasks~\cite{roziere_code_2024, fried_incoder_2023, chen_teaching_2023, joshi_repair_2023} and have already been applied to program synthesis~\cite{gulwani2017program}, \ie, automatically generating programs that accurately correspond to user intents.


As CLLM has been widely used to offer promising solutions for a wide range of coding problems, the quality of the code generated by CLLM has gradually raised concern.
To evaluate the code generation capability of CLLM, existing works have relied on a few small manually curated benchmarks, such as HumanEval~\cite{chen2021evaluating}. 
A coding problem in such benchmarks typically consists of two key components: one input prompt that describes the requirements (\ie, \textit{semantics}) of the problem and the corresponding unit test cases. 
The input prompt provides a {natural language} description of the coding problem, including the function signature, problem description and examples, to the CLLM, which then generates potential solutions. The correctness of the generated code is subsequently verified through unit tests.


\begin{figure}[t]
  \centering
  \includegraphics[width=\linewidth]{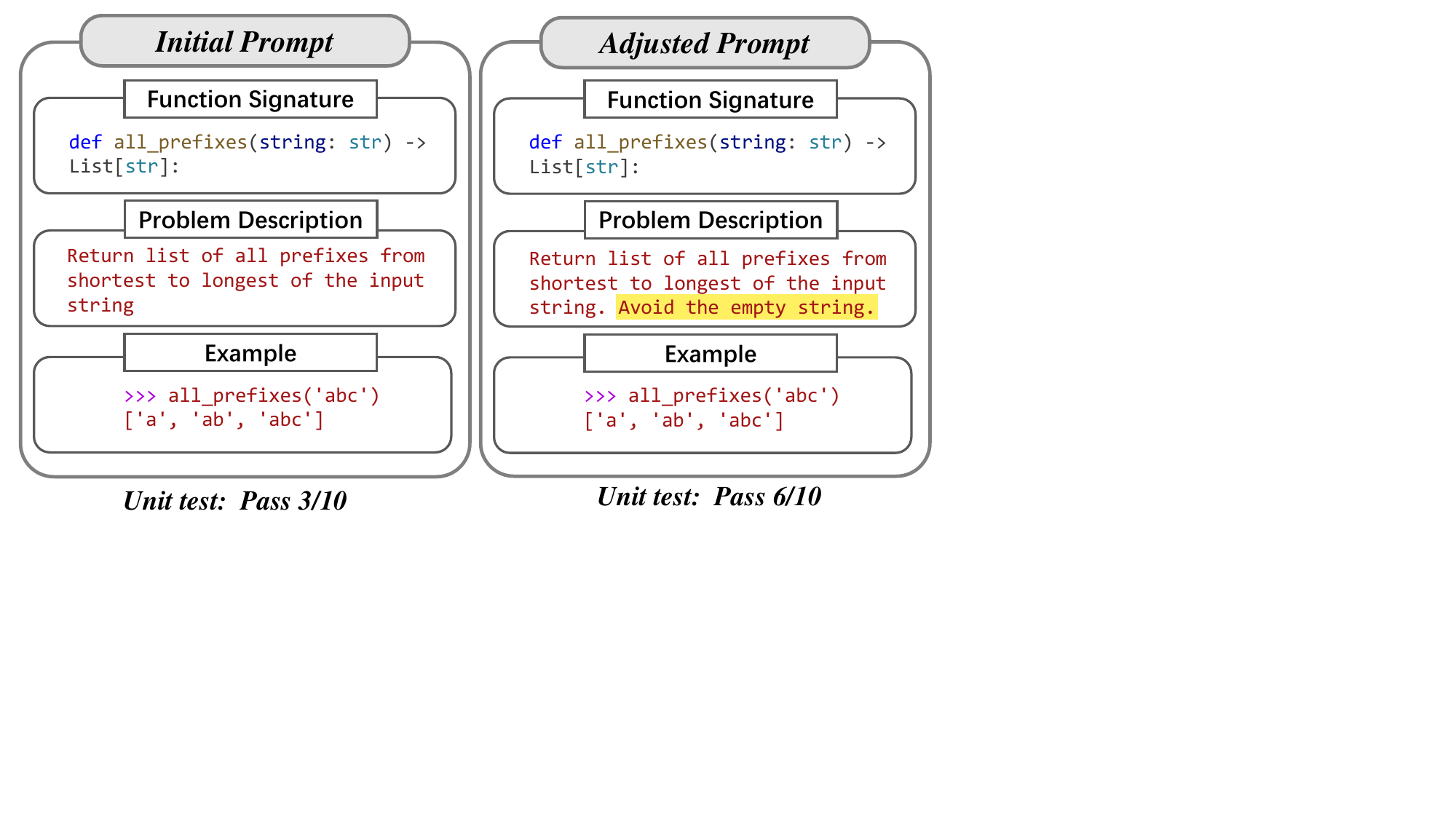}
  \caption{Two input prompts for the same synthesis problem. The difference between the two prompts is highlighted. The correctness of generated code can be significantly improved by making slight adjustments}
  \label{figure:codebench}
\end{figure}

However, there is a substantial gap between real-world scenarios and benchmark settings. Typically, each problem in existing benchmarks is presented with a single prompt, leaving it unclear whether this prompt is the most effective for a CLLM’s comprehension. In practice, a given problem can be described in multiple ways (\ie, different prompts conveying the same semantics), and developers often clarify or request alternative descriptions when a requirement description is unclear. Additionally, developers might describe the same coding problem using varied language, introduce typos, or make spelling errors. Consequently, the evaluation results from existing benchmarks may not accurately reflect the true performance of CLLMs in real-world scenarios. For example, some leaderboards~\cite{noauthor_evalplus_nodate, zhuo2024bigcodebench} have been established for code generation tasks. However, given the influence of requirement descriptions, it is unclear whether these leaderboards truly reflect the accurate ranking of models' performance.


Motivated by these findings, we conducted a comprehensive evaluation to reassess existing code generation benchmarks. Our key observation is that even slight adjustments to a given prompt, without altering the semantics of the coding problems, can significantly impact the performance of CLLMs. For example, as shown in Fig.~\ref{figure:codebench}, adding a minor detail like ``avoid the empty string'' to the problem description increases the number of functionally correct solutions from 3 to 6 out of 10 generated codes. Such pitfalls in existing benchmarks can introduce bias in evaluating the performance of CLLMs, potentially leading to the selection of suboptimal models. Furthermore, these inaccuracies in evaluation may misguide future research and development in improving CLLMs.

In this paper, we aim to thoroughly uncover these pitfalls, with the goal of informing and improving future benchmark design. To achieve this, we propose 10 mutation strategies designed to generate diverse input variations and perturbations for the input prompt in a benchmark and 3 metrics used to measure the impact of input variants and perturbations on benchmark evaluation results before and after mutation. Finally, we generate a new benchmark with 12,834 prompt variants and conduct evaluation on 5 CLLMs. Our key findings are as follows:
\begin{itemize}
\item There is a significant discrepancy between the results from existing benchmarks and the benchmarks including diverse perturbations and variations. Evaluations based solely on existing benchmarks can introduce bias.
\item Mutations in the input prompt can have unexpected effects on CLLMs. For example, providing additional examples may be largely ineffective, while typos in variable names can sometimes improve the model's performance.
\item Mutating different parts of the input prompt can generate variants that impact the correctness of code generation differently across various models. These variants can be used to expand existing benchmarks for a fairer evaluation.
\end{itemize}

%% file: section/2-Background_and_Related_Work.tex
\section{Background and Related Works} \label{background}
\subsection{Code Large Language Model}
Code Large Language Models (CLLMs) have demonstrated impressive capabilities across various coding tasks, including code completion~\cite{fried_incoder_2023}, code translation~\cite{chen_teaching_2023}, and code repair~\cite{joshi_repair_2023}. In these tasks, a code LLM takes a sequence of natural-language instructions and a sequence of code statements (i.e., a partial program) as input, and then generates another partial program as output, depending on the specific task. In this work, we focus on the performance of CLLMs in program synthesis~\cite{gulwani2017program}.

\subsection{CLLM Evaluation}
To assess the code generation capabilities of CLLM, existing works typically rely on a few number of small, manually curated benchmarks. OpenAI constructed the HumanEval~\cite{chen2021evaluating}, the first code problem benchmark to evaluate the performance of CLLM. Since then, many works~\cite{cassano2023multipl, lai2023ds} have focused on building benchmarks that include code problems in multiple languages and of varying complexity. 

Most existing benchmarks~\cite{chen2021evaluating, cassano2023multipl, lai2023ds},  use Pass@k for validating the correctness of a generated function. Pass@k accounts for the inherent randomness in content generated by CLLMs and measures their performance by the proportion of correctly functioning code among multiple generated outputs. In addition to Pass@k, several other metrics are also used to evaluate the code generation quality of CLLMs, such as CodeBLEU~\cite{ren2020codebleu} and CodeBERTScore~\cite{zhou2023codebertscore}. 
Using the proposed benchmarks and metrics, some leaderboards have been established for code generation tasks, such as  EvalPlus Leaderboard~\cite{noauthor_evalplus_nodate}
and BigCode Leaderboard~\cite{noauthor_big_nodate}.


However, these benchmarks do not take into account the input variants and perturbations that exist in practical usage of CLLMs.  Therefore, the performance of CLLMs in practical use may differ from existing benchmarks' evaluation results. In this work, we investigate to what extent input variants and perturbations lead to deviations in benchmark evaluation results.

%% file: section/2.5-Preliminary.tex
\section{Preliminary} \label{preliminary}
\begin{figure*}[htbp]
  \centering
  \includegraphics[width=0.9\linewidth]{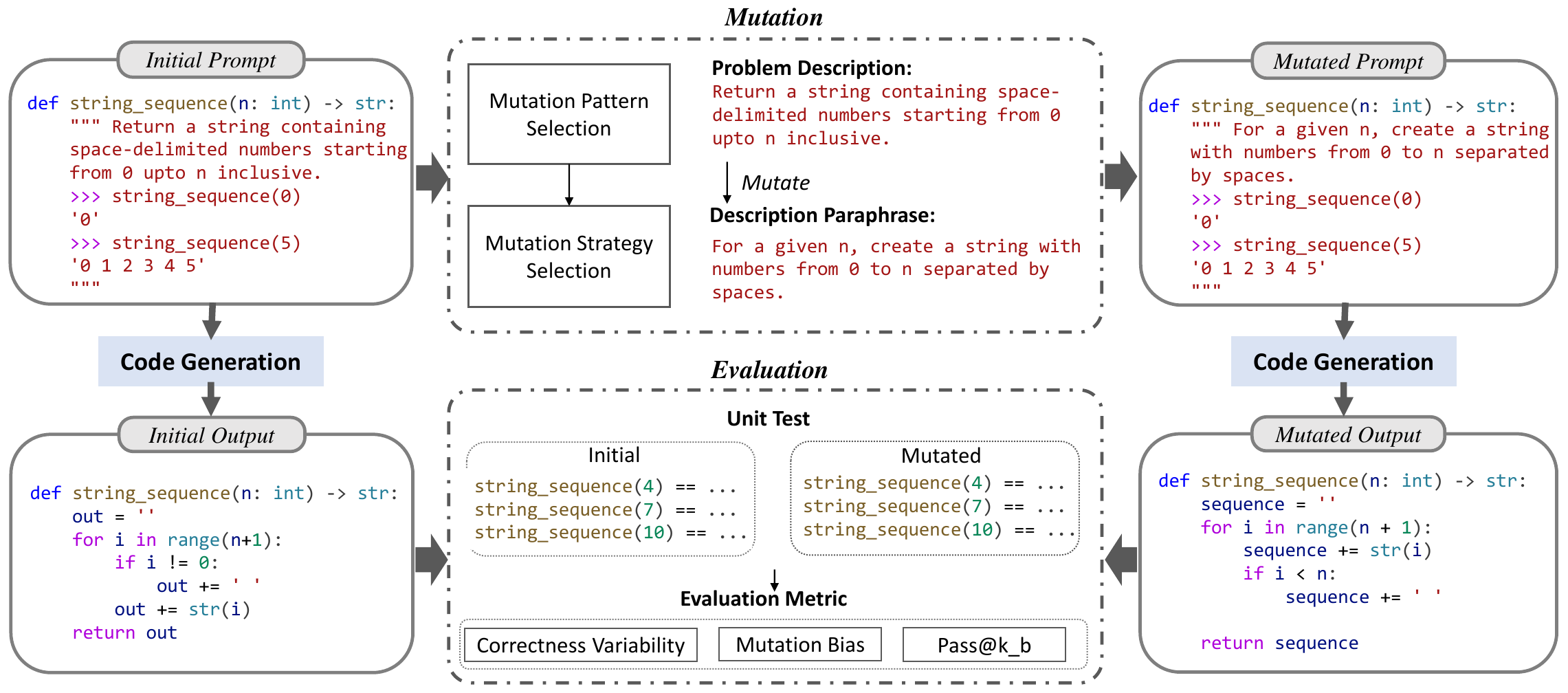}
  \caption{Workflow of our evaluation}
  \label{figure:workflow}
\end{figure*}

\subsection{CLLM Benchmark}
Existing works depend on a limited number of small, manually curated benchmarks to evaluate the performance of CLLMs. As shown in Fig.~\ref{figure:codebench}, for each synthesis problem, the benchmark provides a single input prompt, which consists of three parts:
\begin{itemize}
    \item \textbf{Function Signature:} including the function name and the variable name;
    \item \textbf{Problem Description:} providing a functional description of the synthesis problem in natural language;
    \item \textbf{Example:} providing examples of the inputs and corresponding output results for the synthesis problem.
\end{itemize}

Given an input prompt, a CLLM generates code for solving the coding problem. To facilitate the verification of the correctness of generated code, existing benchmarks~\cite{cassano2023multipl, lai2023ds} provide corresponding unit tests for each synthesis problem, denoted as $UT$, considering code that passes all unit tests as functionally correct. 
We denote the function $pass$ to represent whether the code generated by a model $M$ with the prompt $p$ can pass the unit tests ($UT$), denoted as $pass(M, p, UT)$, where 0 indicates failure and 1 indicates success


In existing benchmarks, there is only one corresponding input prompt for a given synthesis problem. However, in practical usage, even for the same synthesis problem, there may be different ways of describing the problems, \ie, function signature, problem description, and example. Thus, current benchmarks may lack an assessment of CLLMs' ability to solve the same problem with different input variants or a more suitable input. 


\subsection{Pass@k metric}
Pass@k is a commonly used metric to evaluate the correctness of the code generated by CLLMs, which considers the inherent randomness of LLM-generated content and measures how many of the multiple outputs for the same synthesis problem by CLLMs are functionally correct.
In existing works, the correctness of the generated code is determined by unit tests provided by the benchmark. If a generated code can pass all the unit tests, it is considered correct. 
Formally, for each synthesis problem, $n \geq k$ samples are generated, Pass@k counts the number of correct samples $c \leq n$ which pass all unit tests and calculate the unbiased estimator.
{
\small
$$
\operatorname{Pass} \text{@} k:=\underset{\text { Problems }}{\mathbb{E}}\left[1-\frac{\binom{n-c}{k}}{\binom{n}{k}}\right]
$$
}
While Pass@k can account for some randomness in the understanding of inputs, it still measures the success rate based on a single input prompt. This metric does not consider the impact of variations in the input prompts, which could significantly affect the model's performance.


%% file: section/3-Methodology.tex
\section{Methodology} \label{methodology}

\subsection{Overview}
Fig.~\ref{figure:workflow} depicts the workflow of our evaluation. At the mutation step, for each input prompt in a benchmark, a part (\ie, function signature, problem description, or example) of the input prompt will be selected, and a selected mutation strategy will be applied to mutate the selected part to generate a mutated prompt. The initial input prompt and mutated prompts are then fed to the CLLM to obtain the generated code. At the evaluation step, the correctness of the generated code is validated through unit tests.



\begin{figure*}[t]
  \centering
  \includegraphics[width=0.9\linewidth]{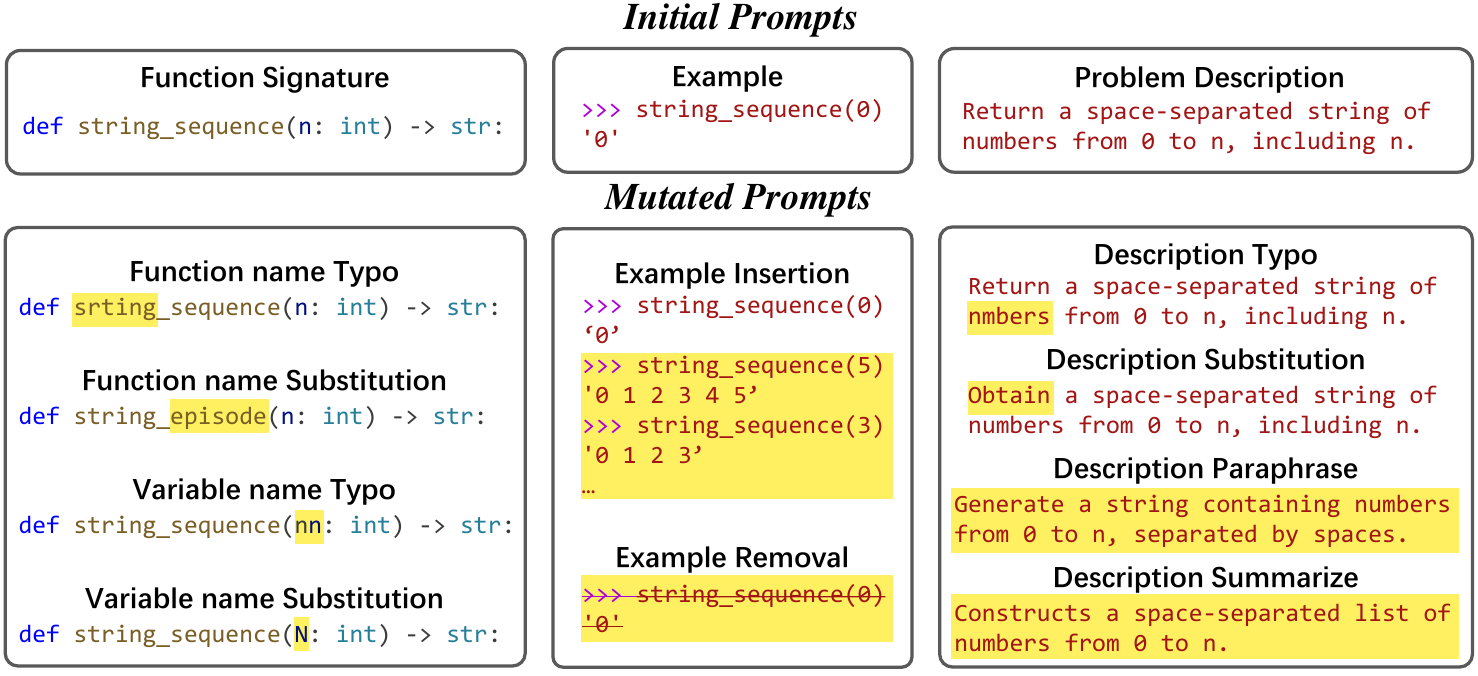}
  \caption{The effects of all mutation strategies, the modifications to the input prompt after mutation are highlighted}
  \label{figure:mutation}
\end{figure*}

\subsection{Mutation Strategy}
As shown in Fig.~\ref{figure:codebench}, the input prompt for each problem includes three parts: function signature, problem description, and examples. To simulate the variants and perturbations that may occur in different parts of input prompts during the practical use of CLLMs, we design 5 transformation methods to mutate each part of the input prompt.
Note that different parts of the input prompt include structured code expressions and natural language expressions, and some transformation methods proposed cannot be applied to both types of expressions simultaneously. 

We will first introduce these five transformation methods and then discuss which transformation methods are applicable to each part of the input prompt.

\subsubsection{Typo Simulation}
This is a character-level transformation method, which simulates typos that may occur during the input process, including:
\begin{itemize}
    \item \textbf{Misspelling}: Simulating possible letter omissions during the spelling process, which does a combination of character-level modifications like delete, insert, and swap. For example, misspelling `close' as `cloes'.
    \item \textbf{Keyboard Typo}: Swapping characters which those close to each other on the keyboard. For example, `threshold' might be misspelled as `theeshold' because the letters `e' and `r' are adjacent on the keyboard.
\end{itemize}
 We utilize the open-source toolbox Augly~\cite{papakipos2022augly} to implement this mutation strategy. The API provided by Augly can simulate the typos that may occur during the input process.

\subsubsection{Synonym Substitution}
This transformation method uses synonyms to replace words in the initial input prompt, changing the expression of the input prompt without altering its semantics. Specifically, for an input prompt, words to be replaced are randomly selected, and a synonym is chosen from WordNet~\cite{miller1995wordnet} to replace the original word.

\subsubsection{LLM Paraphrase}
For the same synthesis problem, there are different ways to describe it. For example, ``Compare the magnitude of two numbers." can also be described as ``Determine which of the two numbers is greater."

This transformation method is designed to generate different problem descriptions for the same synthesis problem. Since LLM has been proven to have exceptional language processing capabilities, we use existing LLM to complete the paraphrase of the initial synthesis problem description. Specifically, we use the following instruction prompt: 
\begin{framed}
    \textit{“I will provide you with a set of prompts. Without changing the meaning of each prompt, rewrite each one in 10 different ways.”} 
\end{framed}
    

\subsubsection{LLM Summarize}
This transformation method aims to enhance the accuracy of the description based on the function signature and examples. We also utilize LLMs to implement this method.
Specifically, for an original input prompt, we extract the problem description and ask the LLM to improve the problem description based on the function signature and the example. The instruct prompt we used is as follows: 
\begin{framed}
    \textit{``I will provide you with ten function signatures and corresponding test cases, along with the description of the function's functionality. Please improve the accuracy of the description to make it more suitable for all test cases of each function. For each function, provide me with ten improved descriptions.''}
\end{framed}

\subsubsection{Example Insertion and Removal} 
This transformation method randomly adds or deletes the number of examples contained in the input prompt, to verify the impact of the number of examples provided by the input prompt on the quality of code generation. 


\subsubsection{Mutation Application} 
Since the function signature and example sections of the input prompt are structured code, while the problem description is in natural language, certain transformation methods cannot be uniformly applied to both types of expressions.

Specifically, for the function signature, we apply \textit{Typo} and \textit{Synonym Substitutions} to function and variable names. For the problem description, we apply \textit{Typos}, \textit{Synonym Substitutions}, \textit{LLM Paraphrase}, and \textit{LLM Summarize}. For the examples, we mutate them by adding or removing instances.

The mutation strategies applied to each part, the impact of mutation, and examples can be seen in Fig.~\ref{figure:mutation}.

\subsection{Evaluation Metric}
Pass@k, the most widely used metric, cannot be used to measure changes in the correctness of output code after input mutations because it only works for a single input prompt. Therefore, in order to quantify the changes in benchmarks with diverse mutations, we propose 3 metrics: \textit{Correctness Variability} (CV), \textit{Mutation Bias} (MB) and \textit{Best Pass@k} ({Pass@k\_b}).
\begin{itemize}
    \item \textbf{Correctness Variability (CV):} Given a mutated prompt, CV measures the differences in terms of the number of correct code generated between it and its original prompt. Given a CLLM $M$, for a mutated prompt $p_{m}$ as well as the corresponding initial input prompt $p_{ini}$, 
    $n$ code samples are generated by $M$, respectively.  
     CV calculates the difference between the number of correct samples generated using $p_{m}$ and $p_{ini}$. Formally, CV for a single prompt is expressed as: 
    {
    \small
    $$
    \mathrm{CV}_{\mathrm{n}}\left(P_m\right)=\sum_n \operatorname{pass}\left(M, p_m, U T\right)-\sum_n \operatorname{pass}\left(M, p_{i n i}, U T\right)
    $$
    }
    \item \textbf{Mutation Bias (MB):} MB measures the impact of a set of input variants generated through mutation on the correctness of CLLM-generated code. Given a set of input variants $D$ generated through mutation, MB first calculates the Correctness Variability for all input prompts in the dataset. To measure the overall impact of input variants on CLLM-generated code, MB then calculates the average of the absolute values of all Correctness Variabilitys. Formally, MB is expressed as:
    {\small
    $$
    \operatorname{MB}(\mathrm{D}, \mathrm{M})=\frac{\sum_{p \in D}\left|CV_n(p)\right|}{|D|}
    $$}
    
    \item \textbf{Pass@k\_b:} Pass@k\_b calculates the optimal Pass@k value for a set of input variants. Given a set of input variants $D$ generated through mutation, Pass@k\_b first computes the Pass@k for each variant and then takes the maximum value among them. Formally, Pass@k\_b is expressed as:
{\small
$$
\operatorname{Pass} @ \mathrm{k}\_b=\max \left\{ \operatorname{Pass} @ \mathrm{k}(p) | p\in \mathrm{D} \right\}$$}
where $\operatorname{Pass} @ \mathrm{k}(p)$ represents the Pass@k results with the prompt $p$.


\end{itemize}




%% file: section/4-Experiments.tex
\section{Experiments} \label{experiments}
We have implemented our evaluation framework using Python 3.8, and all experiments were conducted on an NVIDIA A800. To evaluate the performance of CLLMs with our new benchmarks, we aim to answer the following research questions (RQs):

\begin{table}
\centering
\caption{Number of variants generated by each mutation strategy}
\small
\label{tab:number}
\begin{tabular}{cc}
\hline 
Mutation Strategy & Variants Number\tabularnewline
\hline 
Function Name Typo & 1629\tabularnewline
Function Name Substitution & 1093\tabularnewline
Variable Name Typo & 1640\tabularnewline
Variable Name Substitution & 740\tabularnewline
Description Typo & 1640\tabularnewline
Description Substitution & 1636\tabularnewline
Description Paraphrase & 1640\tabularnewline
Description Summarize & 1640\tabularnewline
Example Insertion & 1012\tabularnewline
Example Removal & 164\tabularnewline
Total & 12834\tabularnewline
\hline 
\end{tabular}

\end{table}

\begin{table*}[t]
\centering
\caption{Results of MB, Pass@1, and Pass@1\_b of different models under the influence of various mutation strategies.}
\label{tab:main}
\small
\begin{tabular}{ccccccc}
\hline 
Mutation Strategy & Metric & DeepSeek & Llama3.1 & CodeLlama & CodeGen & InCoder\tabularnewline
\hline 
Initial Prompt & Pass@1 & 67.99 & 57.56 & 22.0 & 21.8 & 7.87\tabularnewline
\hline 
\multirow{3}{*}{Function Typo} & MB & 10.46 & 10.41 & 7.40 & 7.22 & 3.21\tabularnewline
 & Pass@1 & 67.24 & 57.40 & 19.13 & 19.59 & 6.33\tabularnewline
 & Pass@1\_b & 79.02 & 69.82 & 28.41 & 29.02 & 10.00\tabularnewline
\hline 
\multirow{3}{*}{Function Substitution } & MB & 12.26 & 10.94 & 7.02 & 7.99 & 2.91\tabularnewline
 & Pass@1 & 64.78 & 55.12 & 18.06 & 17.79 & 5.58\tabularnewline
 & Pass@1\_b & 78.91 & 69.29 & 26.62 & 27.43 & 8.38\tabularnewline
\hline 
\multirow{3}{*}{Variable Typo } & MB & 8.82 & 8.95 & 3.98 & 3.97 & 1.83\tabularnewline
 & Pass@1 & 68.87 & 58.52 & 20.02 & 20.36 & 6.98\tabularnewline
 & Pass@1\_b & 79.08 & 69.62 & 24.94 & 25.30 & 9.39\tabularnewline
\hline 
\multirow{3}{*}{Variable Substitution} & MB & 8.82 & 8.66 & 12.66 & 12.07 & 5.56\tabularnewline
 & Pass@1 & 70.09 & 59.78 & 16.22  & 16.34  & 4.54\tabularnewline
 & Pass@1\_b & 79.01 & 69.50 & 25.55 & 26.80 & 9.53\tabularnewline
\hline 
\multirow{3}{*}{Description Typo } & MB & 11.20 & 11.23 & 6.37  & 7.44  & 2.20\tabularnewline
 & Pass@1 & 65.41 & 57.49 & 18.85  & 18.74  & 7.10\tabularnewline
 & Pass@1\_b & 78.76 & 69.39 & 23.54 & 25.00 & 9.57\tabularnewline
\hline 
\multirow{3}{*}{Description Substitution } & MB & 11.22 & 10.72 & 7.14  & 7.73  & 2.85\tabularnewline
 & Pass@1 & 65.24 & 57.56 & 17.72  & 18.29  & 6.33\tabularnewline
 & Pass@1\_b & 78.62 & 69.33 & 24.27 & 26.04 & 10.00\tabularnewline
\hline 
\multirow{3}{*}{Description Paraphrase } & MB & 16.69 & 14.90 & 8.12 & 8.30 & 2.53\tabularnewline
 & Pass@1 & 60.13 & 54.12 & 18.96 & 19.28 & 6.33\tabularnewline
 & Pass@1\_b & 78.00 & 69.27 & 26.95 & 28.23 & 10.73\tabularnewline
\hline 
\multirow{3}{*}{Description Summarize} & MB & 18.32 & 16.34 & 11.18 & 9.18 & 2,69\tabularnewline
 & Pass@1 & 60.20 & 54.48 & 41.65 & 19.19 & 6.85\tabularnewline
 & Pass@1\_b & 77.92 & 69.38 & 53.50 & 29.76 & 10.73\tabularnewline
\hline 
\multirow{3}{*}{Example Insertion} & MB & 12.12 & 11.33 & 5.95 & 5.80 & 1.97\tabularnewline
 & Pass@1 & 62.53 & 50.35 & 14.71 & 14.88 & 4.53\tabularnewline
 & Pass@1\_b & 77.61 & 68.89 & 24.68 & 24.75 & 8.48\tabularnewline
\hline 
\multirow{3}{*}{Example Removal} & MB & 17.13 & 10.34 & 9.51 & 8.60 & 3.11\tabularnewline
 & Pass@1 & 56.34 & 54.94 & 18.84 & 20.18 & 7.20\tabularnewline
 & Pass@1\_b & 56.34 & 54.94 & 18.84 & 20.18 & 7.20\tabularnewline
\hline 
\multirow{3}{*}{Average} & MB & 12.80 & 11.46 & 7.61 & 7.73 & 2.82\tabularnewline
 & Pass@1 & 64.44 & 56.54 & 19.68 & 19.10 & 6.60\tabularnewline
 & Pass@1\_b & 75.40 & 67.44 & 25.99 & 26.23 & 9.41\tabularnewline
\hline 
\end{tabular}

\end{table*}

\begin{itemize}
    \item \textbf{RQ1:} Can the existing benchmarks and metrics fairly reflect the performance of CLLMs?
    \item \textbf{RQ2:} How does the input variants and perturbations affect the output of CLLMs?
    
\end{itemize}
\subsection{Setup}
\subsubsection{Datasets}
Following the previous works~\cite{min2023beyond, hooda2024large, allamanis2024unsupervised}, we conduct our evaluation using the widely-used synthesis problem benchmark: HumanEval. Humaneval is constructed by OpenAI to evaluate the code generation capabilities of large language models, which consists of 164 handwritten synthesis problems. Each problem includes a function signature, problem descriptions, examples, and scripts for execution evaluation.


\subsubsection{Models}
We conduct our evaluation on 5 popular Code Generation Models that are commonly used in existing works:
DeepSeek (6.7B)~\cite{guo2024deepseek}, Llama3.1 (8B)~\cite{dubey2024llama},
CodeLlama (7B)~\cite{roziere_code_2024}, CodeGen (7B)~\cite{nijkamp2022CodeGen}, InCoder (7B)~\cite{fried2022incoder}. For these CLLMs, we use the same temperature, which is set as 0.8.


\subsubsection{Mutation Configuration}
For the 6 mutation strategies using \textit{Typo} and \textit{Synonym Substitution}, we attempted to generate 10 variants for each input prompt using each mutation strategy. However, since some input prompts contain function names and variable names that may not be standard words, making it hard to create 10 variants (\eg, finding 10 synonyms or typos), we generated the maximum number of variants possible for those prompts.

For \textit{Description Paraphrase} and \textit{Description Summarize}, we use GPT-4o~\cite{noauthor_hello_nodate} with the instruction prompts proposed to generate 10 variants for each input prompt.
For\textit{ Example Insertion}, we extracted the input-output pairs from the corresponding unit tests and inserted them into the input prompt following the example format in the prompt. We attempted to generate 10 variants for each input prompt by adding 1, 2, ..., 10 examples. However, due to the lack of sufficient samples in some unit tests, we generated the maximum number of variants possible for those cases. For \textit{Example Removal}, we removed all examples included in each input prompt.

The number of variants generated by each mutation strategy is shown in Table~\ref{tab:number}. In total, we generated 12834 variants for our evaluation.

\subsubsection{Experimental Setup}

We use all the variants generated by all 10 mutation strategies. For each variant, we generate code 10 times using each model. As CLLM often does not automatically stop after generating a complete function, we extract the first complete function from each generated output for subsequent unit testing and analyze the impact of mutations on benchmark evaluation results.

For \textbf{RQ1}, we use CLLM to generate code for all input variations and evaluate its performance using three metrics: the average Pass@1, the average Pass@1\_b,
and the Mutation Bias (MB) for all variations generated by each mutation strategy.
For \textbf{RQ2}, we analyze the impact of different mutation strategies on the generated code. Specifically, we calculate and compare the average MB of variants generated by different mutation strategies.

\subsection{Results}

\begin{figure*}[t]
  \centering
  \includegraphics[width=0.85\linewidth]{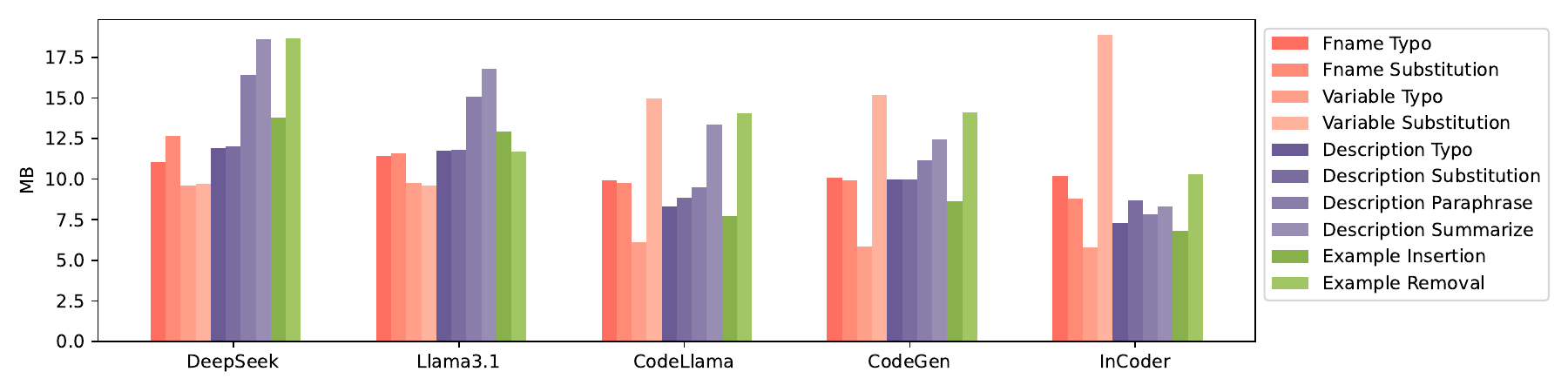}
  \caption{MB across Mutation Strategies} 
  \label{figure:mutation_boxplot}
\end{figure*}
\subsubsection{\textbf{RQ1:}}

Table~\ref{tab:main} shows the Pass@1, Pass@1\_b, and Mutation Bias (MB) for the five models.

\textit{Finding 1: There is a discrepancy between CLLM performance and benchmark evaluation results in practical scenarios.} For all models, MB is a non-zero value, indicating that various mutations will impact the correctness of code generated by CLLMs.
In real-world usage scenarios with variants and perturbations, there is a difference between the performance of CLLMs and the performance results validated through benchmarks.




\textit{Finding 2: Existing benchmarks do not fairly reflect the performance of CLLMs due to the limited problem description.  Different kinds of descriptions for the same synthetic problem can significantly alter the evaluation results of the current benchmarks.} Comparing the Pass@1 results of each model on the initial prompt with the Pass@1\_b results under various mutation strategies, we find that Pass@1\_b consistently outperforms the initial prompt's Pass@1. 
This indicates that existing benchmarks only reflect the performance of a synthesis problem under a specific description, which is not a fair assessment comparison of CLLM performance. Given that CLLMs may prefer certain problem descriptions over others, there could be alternative descriptions that yield better outputs for each CLLM. 
Comparing the results between CodeLlama and CodeGen, CodeLlama outperforms CodeGen on the initial prompt (\ie, Pass@1: 22.0 vs. 21.8). However, in the presence of input variations and perturbations, the average Pass@k\_b of CodeGen is higher than that of CodeLlama (26.23 vs. 25.99). 

\textit{Finding 3: CLLMs that perform better on existing benchmarks tend to show greater discrepancies between their performance under input variants and disturbances compared to benchmark evaluation results.} Compare the Pass@k and MB of different models. We also find that a model's strong performance on Pass@k does not necessarily indicate that it performs better when faced with input variants and perturbations. On the contrary, in most cases, models that perform better in Pass@1 tend to exhibit greater fluctuations in performance when faced with input variants and perturbations. For example, comparing DeepSeek and CodeLlama, while DeepSeek has a better Pass@k (67.99 vs 22.0), CodeLlama shows a smaller MB under the influence of most mutation strategies (9 out of 10). 






\subsubsection{RQ2:}
Fig.~\ref{figure:mutation_boxplot} shows the impact of variants generated by different mutation strategies on the accuracy of code generation for each model. 

\textit{Finding 4: For different models, the mutation of different parts of the input prompt affects the quality of the generated code differently.}
Overall, the impact of mutations in different parts of the input prompt varies, and this effect also differs between models. For example, for Llama3.1, mutations in the problem description have a greater impact on the correctness of the generated code, while for InCoder, mutations in the function name section have a more significant effect.

\textit{Finding 5: The variable names in the input prompt have a stronger impact on the code generation accuracy of lower-performing models. Typos in variable names can sometimes improve the model's evaluation performance.}
Regarding the function signature, we find that CLLMs with relatively lower performance (refer to Table~\ref{tab:main}) are more susceptible to the impact of \textit{Synonym Substitutions} for variable names.
This indicates that these models are more influenced by variable names when generating code and have relatively weaker natural language understanding, as they cannot fully comprehend synonyms.

Moreover, according to the results of function signature mutations in Table 1, in most cases, mutations to variables lead to a decrease in Pass@1. However, for DeepSeek and Llama3.1, the two types of variable mutations actually resulted in an increase in Pass@1. For example, under the influence of Variable Typo, DeepSeek's Pass@1 is 68.87, which is higher than the results on initial prompt (\ie, 67.99).



\textit{Finding 6: Mutations in the problem description of the input prompt have a greater impact on CLLM with high performance.}
For the problem description, we find that for CLLMs with better performance (as shown by Pass@1 in Table~\ref{tab:main}), mutations in the problem description of the input prompt have a greater impact on the correctness of code generation. This indicates that for high-performing models, the way the problem description is phrased and its accuracy are crucially important. This further suggests that the current benchmark, which provides only a single input prompt for each synthesis problem, may lead to unfair comparisons of CLLMs' performance, particularly when evaluating higher-performing models.


\textit{Finding 7: Providing more input-output examples in the input prompt does not necessarily improve the correctness of code generated by CLLMs.}
Regarding the examples, we found that removing the example has a greater impact on the correctness of generated code on DeepSeek, CodeLlama, CodeGen, and InCoder, and smilar impact to Example Insertion on Llama3.1. Moreover, comparing the Pass@1 results of example insertion and example removal, we find that for Llama 3.1, CodeLlama, CodeGen, and InCoder, the correctness of code generation is higher when the examples are removed than when more examples are added. This suggests that these four models do not effectively utilize the information from the examples for code generation. In contrast, DeepSeek is able to leverage additional examples to improve code generation accuracy.

%% file: section/5-Discussion.tex
\section{Discussion} \label{discussion}

\textbf{Impact of Mutation Variability on Benchmark Bias and Model Robustness}:
Our findings reveal a significant variance in model performance when subjected to different types of prompt mutations. This variability underscores the risk of relying on a single prompt for evaluation, as different models may have distinct preferences and strengths in interpreting the provided prompts. Such discrepancies can introduce bias in the evaluation process, potentially leading to an inaccurate assessment of a model's true capabilities. Therefore, we recommend that future evaluations incorporate a broader range of prompts to assess model robustness comprehensively. Metrics like Pass@k\_b and MB might offer a more accurate reflection of this impact, as they allow for the comparison of models based on their optimal performance across a variety of suitably tailored prompts, providing a fairer and more informative evaluation.

\textbf{Enhancing Mutation Strategies for Comprehensive Model Evaluation}:
While our study introduces several mutation strategies to evaluate code generation models, there is still room for improvement by incorporating more diverse mutations. Given the varying capabilities of models in understanding different prompts, it is crucial to ensure that models truly comprehend the problem before generating code and the comparisons. To achieve this, an iteration-based evaluation approach could be implemented, where models are allowed to generate code only after confirming their understanding through multiple clarification iterations. This process would involve refining the prompts and addressing any unclear aspects iteratively, ensuring that the final code generation is based on a clear and accurate understanding of the problem. Such an approach could further mitigate biases and lead to a more reliable assessment of a model's true capabilities in real-world coding scenarios.

%% file: section/6-Conclusion.tex
\section{Conclusion} \label{conclusion}

In this paper, we explore the pitfalls in existing code generation benchmarks. We access five popular CLLMs using 12,834 generated prompt variants, and found a significant performance discrepancy between the results from existing benchmarks and those from mutated benchmarks containing perturbations and variations. This finding underscores the need for more robust evaluation methods and benchmarks.
